\documentclass[twocolumn,fleqn]{svjour3}    % twocolumn
\smartqed  % flush right qed marks, e.g. at end of proof
\usepackage{graphicx}
 \usepackage{mathptmx}      % use Times fonts if available on your TeX system
 \journalname{Low Temperature Physics}
\begin{document}

\title{ Quantitative semiclassical  analysis of ultracold weakly interacting   bose gas trapped in optical boxes}


\author{ Ahmed S. Hassan, Azza M. Elbadry,  A. M. Mohammedein}


\institute{Ahmed S. Hassan, Azza M. Elbadry\at
               Department of Physics, Faculty of Science,  Minia  University, El Minia, Egypt.\\
              Tel.: +201006592341\ ,
              Fax:  0020862363011\\
              \email{ahmedhassan117@yahoo.com}            %  \\
							\and 
							A. M. Mohammedein \at
							Department of Laboratory Technology, College of Technological Studies, PAAET PO Box 42325, Shuwaikh 70654, Kuwait.
						} 

\date{Received: date / Accepted: date}
% The correct dates will be ente  dotted red by the editor

% 

\maketitle

\begin{abstract}
In this paper, the condensate fraction and the critical atom number and its corresponding critical temperature of condensate ultracold   boson atoms trapped in optical box traps,  are investigated. The semiclassical approximation is employed in this study.  The boxes traps are  modeled by a general power-law potential.  The  deviation of the  boxes traps  from an ideal boxes traps are discussed.
The out come results furnish  useful quantitative theoretical results for the
future BEC experiments in such traps.
\end{abstract}

%\PACS{ 03.75.Kk \and 37.10.Vz \and 67.85.-d \sep 67.90.+z }    % PACS, the Physics and Astronomy

\section{ Introduction}
Recently a wonderful development has been in the demonstration of BEC or BEC-like transition in ultracold alkali gases trapped in optical boxes \cite{Hadzibabic,Hadzibabic1}.  An interesting result of this research concerns the role of the external trapping potential. 
The form of the trapping potential is strongly related to the macroscopic behavior of the condensate.
A common way to characterize (imperfect) box traps is to model them by a power-law potential isotropic  \cite{Gaunt}. 
A lot of efforts are made to measure the typical values for the number of atoms and temperature at the BEC transition are measured. These values are known as the critical points , $(N_c, T_c)$. These critical points are corresponding to the threshold of BEC. The measured data are deviated from both the ideal gas and the calculated results based on the mean field theory \cite{dav,Shi,Zobay}.  

\par   
In this work we want to provide an approach which enables one to deal with any kind of power-law potential conditions, including the anisotropic harmonic oscillator potential and the most relevant rigid boxes potential. Our motivation  is  to extend and clarify the previous works, and also provide a more
detailed discussion of the physical contents of the theory \cite{Bagnato1,Bagnato2,klaus1,Celia,Salanich,Suzuki,Faruk}.  In spite of the interest in this topic and the progress made so far, none of these works provides a complete picture of the problem. The goal of this work is to fill this gap.

The methods we are going to employ is the semiclassical approximation \cite{gro,kir,ket1,hau,pat}, which has been   
employed when considering the properties of Bose-condensed ideal gases trapped in power-law potentials.
 We have calculated the grand canonical potential relevant to the power-law potential. 
Employing partial derivative of the grand potential, an analytical expression for the  condensate atoms number $N_0$ in the ground state, and the number of atoms in the excited state $N_{th}$ are calculated. Both of them are used to
 calculate the condensed fraction, and the critical atoms number and its corresponding critical temperature.
The corrections due to the interatomic interaction are discussed explicitly \cite{Gio1,Gio2,Gio3}. The results
obtained here give not only many significant conclusions in literature but also some new characteristics about
the trapped interacting Bose gases. So it can be used to describe the thermodynamic properties of a class of
interacting as well as non-interacting Bose gases in a unified way.

\par 
The  paper is planed as follows: Section two includes a simple model for the BEC
in 3D power-law potential. Section three is devoted to investigate our semiclassical approximation. Section four is devoted to calculate the condensed fraction and the critical points.
Section five presents a short discussion and conclusion. 
%========================================================================   
\section{ Physical model}
 In this section, we introduce the necessary theoretical basis and the nomenclature relevant to the present work.
A common potential to characterize the box traps is to
model them by an isotropic power-law potential $V({\bf r}) \propto {\bf r}^p$, with $p \to \infty$.  We thus
consider a degenerate Bose gas trapped in a power-law potential given by
\begin{equation}
 V({\bf r}) = \epsilon_1\ \big|\frac{x}{a}\big|^p + \epsilon_2\ \big|\frac{y}{b}\big|^l + \epsilon_3\ \big|\frac{z}{c}\big|^q,  
 \label{eq1}
\end{equation}

The Hamiltonian describing the interacting atomic gas in the potential (\ref{eq1}) is given by \cite{Bagnato1}
\begin{eqnarray}
H({\bf r,p}) &=& \frac{ p^2}{2m} + V({\bf r}) + 2 g n(\bf r), \nonumber\\
             &\cong& \frac{ p^2}{2m} + V_{eff}({\bf r})  
\label{eq2}
\end{eqnarray}
with $p^2 = p_x^2 + p_y^2 + p_z^2$ is the momentum, $m$ is the atom mass, $g = \frac{4\pi\hbar^2 a }{m}$ is the interaction strength with $a$ is the s-wave scattering length
and $V_{eff}({\bf r}) $ is the effective potential,
\begin{equation}
	V_{eff}({\bf r})  = V({\bf r})  + 2 g [n_{th} (x, y, z) + n_0(x, y, z)],     
	\label{eq3} 
\end{equation}
  
%  here    

Usually,  BEC is described within the grand canonical ensemble. All relevant thermodynamic quantities can be calculated from partial derivative of the grand potential $q(T)$, which is the logarithm of the grand canonical partition function \cite{kir}.
\begin{equation}
q(T) = - \sum_{n=0}^\infty \ln (1 - e^{-\beta (E_n- \mu)})
\label{eq-4}
\end{equation}  
where  $E_n$ is the eigenvalues for the potential  Eq.(\ref{eq2}) and $\mu$ is the chemical potential of the  condensate boson. 
 It is convenient to separate out the ground state contribution and expand the logarithm, $\ln(1-y)  = - \sum_{j=1}^\infty \frac{y^j}{j}$, to express $q$ as a sum over
Bose-Einstein distribution\cite{ket1}, 
$
	N_n = \frac{{\textsc z} e^{-\beta E_n}}{1 - {\textsc z} e^{-\beta E_n}} = \sum_{j=1}^\infty {\textsc z}^j \sum_{n=0}^\infty e^{-j\beta E_n}.
	$
Thus, Eq.(\ref{eq-4}) can be rewritten as,
\begin{eqnarray}
q(T) &=& q_{0} + \sum_{j}\frac{{\textsc z}^j}{j} \sum_{n=1}^\infty e^{-j\beta  E_n }  \nonumber\\    
&\equiv&  q_0 + q_{th}
\label{eq4-1}
\end{eqnarray}
with $ q_{0}=-\ln (1-  {\textsc z}) $ is the grand potential for the atoms in the ground state, $q_{th}$ is the grand potential for thermal atoms and  $ {\textsc z} = e^{\beta (\mu - E_0)}$ is the effective fugacity.  
 
\section{Semiclassical approximation}  
The sum in Eq.(\ref{eq4-1}) cannot be evaluated analytically in a closed form. Another possible way to do
this analysis,  the sum over $n$ in Eq.(\ref{eq4-1}) can be converted into an integral over the phase space by replacing the discrete $E_n$ with a continuous variable $\epsilon({\bf r; p})$ depending on  position $\bf r$ and momentum $\bf p$, which corresponds to the classical energy associated with the single-particle Hamiltonian for the system given in Eq.(\ref{eq2}) \cite{Celia,hau}.   Within this approximation Eq.(\ref{eq4-1}) becomes

\begin{eqnarray}   
	q_{th} ({\bf p, r}) &=& -\frac{1}{(2\pi \hbar)^3} \sum_{j=1}^\infty \frac{{\textsc z}^j}{j}  \int e^{-j\beta[\frac{p^2}{2m} +  V_{eff}({\bf r})]}   { d{\bf p} d{\bf r}}
\label{eq5} 
\end{eqnarray}  
After doing the $p$ integration,  the local grand potential is given by   
\begin{equation}
q_{th} ({\bf  r}) = \frac{1}{\lambda_{th}^3}   \sum_{j=1}^\infty \frac{{\textsc z}^j }{j^{5/2}} \int  e^{- j\beta  V_{eff}({\bf r}) } d{\bf r} 
	\label{eq5-2} 
\end{equation}
where $\lambda_{th} = \sqrt{\frac{2 \pi \hbar^2}{m k_B T}}$ is the thermal de-Broglie wavelength. However, calculating the phase space integral required calculating  the densities of condensate, thermal atoms and the chemical potetial. The above mentioned parameters are calculated using the Hartree-Fock approximation and given in the next subsection.  

In order to calculate the integral given in  Eq.(\ref{eq5-2}), we follow  the Hadzibabic and co-worker \cite{tammuz,hadz1,hadz2,hadz3,hadz4} approach's and consider the same approximation \cite{has2,has3,has4}.  
Within this approximation  the thermal component is treated as a gas of non-interacting atoms which are moving in the effective potential given in Eq.(\ref{eq3}).
Further simplifications is made as a consequence of the relative diluteness of the thermal component compared to the condensate component.
 If the effect of thermal atoms on the condensate is neglected the condensate component $n_0(\bf{r})$ is given by the
Thomas-Fermi approximation for the time independent Gross-Pitaevskii equation which describe the condensate atoms part. If the mean-field energy of the thermal atoms $2gn_{tr}(\bf{r})$ is also neglected the resulting effective potential experienced by the thermal
atoms is  given by
\begin{eqnarray}
	V_{eff}({\bf r}) &=& V({\bf r})  + 2 g  n_0({\bf r}), \nonumber\\  
	       &=& | V({\bf r})  - \mu| + \mu 
				\label{eq14}
\end{eqnarray}  
by substituting from Eq.({\ref{eq14}}) in Eq.({\ref{eq5-2}}) leads to
\begin{eqnarray}
 q_{th} ({\bf  r}) &=&  \frac{1}{\lambda_{th}^3}\sum_{j=1}^\infty \frac{1 }{j^{5/2}} \int  e^{-j\beta\big(\epsilon_1\ \big|\frac{x}{a}\big|^p + \epsilon_2\ \big|\frac{y}{b}\big|^l + \epsilon_3\ \big|\frac{z}{c}\big|^q - \mu \big)}  d{\bf r}\nonumber\\
&=&  \frac{1}{\lambda_{th}^3}\sum_{j=1}^\infty \frac{1 }{j^{5/2}} \int  e^{-j\big( \big|\frac{x}{R_x}\big|^p +  \big|\frac{y}{R_y}\big|^l +  \big|\frac{z}{R_z}\big|^q - \alpha_0\big)}  {dxdydz}\nonumber\\
&=&  \frac{1}{\lambda_{th}^3}\sum_{j=1}^\infty \frac{1 }{j^{5/2}} \int_0^\infty  e^{-j\big( \big|\frac{x}{R_x}\big|^p\big)} dx \int_0^\infty  e^{-j\big(\big|\frac{y}{R_y}\big|^l\big)} dy \nonumber\\ 
&\times& \int_{R_x\alpha_0^{\frac{1}{q}}}^\infty  e^{-j\big(\big|\frac{z}{R_z}\big|^q - \alpha_0\big)}  {dz}
	\label{eq152} 
\end{eqnarray}   
where $R_x(T),R_y(T)$ and $R_z(T)$ are the
 thermal radii, equivalent to the Thomas-Fermi radii, which fixed the maximum value of the chemical potential compared to $k_B T$, 
\begin{eqnarray}
R_x(T) &=& \Big(\frac{a^p}{\epsilon_1 \beta}\Big)^\frac{1}{p},\ 
R_y(T) = \Big(\frac{b^l}{\epsilon_2 \beta}\Big)^\frac{1}{l},\ R_z(T) = \big(\frac{c^q}{\epsilon_3 \beta}\Big)^\frac{1}{q}\nonumber\\
\alpha_0 &=& \frac{\mu}{k_BT}
\end{eqnarray}
these radii are  equivalent to  the condensate Thomas-Fermi radii at  which the thermal density
drops to zero along $T \to 0$. 

Now it is straightforward to calculate the integrals in Eq.(\ref{eq152}) using   the definition of the gamma function. Using the definition of the gamma function $\Gamma(s) =\int_0^\infty t^{s-1} e^{-t} dt$ we have, 
\begin{equation}
	\int_0^\infty   e^{-j\big( \big|\frac{x}{R_x}\big|^p \big)} dx = \frac{R_z(T)}{p\ j^{1/p}} \Gamma(\frac{1}{p})
		\label{eq1531} 
\end{equation}
and
\begin{equation}
	\int_0^\infty   e^{-j\big( \big|\frac{y}{R_y}\big|^l \big)} dy = \frac{R_y(T)}{p\ j^{1/l}} \Gamma(\frac{1}{l})
		\label{eq153} 
\end{equation}
where $t = j \big(\frac{x}{R_x}\big)^p$ and $t = j \big(\frac{y}{R_y}\big)^l$ is used here.
For the $z$-integral,
setting $t = j\big( \big|\frac{z}{R_z}\big|^q - \alpha_0\big)$ leads to,
\begin{eqnarray}
 \int_{R_x\alpha_0^{\frac{1}{q}}}^\infty  e^{- j\big( \big|\frac{z}{R_z}\big|^q - \alpha_0 \big) }dz &=&\frac{R_z}{qj^{\frac{1}{q}}} \int_0^\infty t^{\frac{1}{q}-1} \big(1 + \frac{j}{t} \alpha_0 \big)^{\frac{1}{q}-1} e^{-t} {dt}\nonumber\\
&\approx&  \frac{R_z}{qj^{\frac{1}{q}}} \int_0^\infty t^{\frac{1}{q}-1}  \big(1 + \frac{j(\frac{1}{q}-1 )}{t} \alpha_0 \big) e^{-t} {dt}\nonumber\\
&\approx& \frac{R_z}{qj^{\frac{1}{q}}}\Big[  \Gamma(\frac{1}{q}) + (\frac{1}{q}-1 ) \Gamma(\frac{1}{q}-1) j\ \alpha_0 \Big] \nonumber\\
&\approx& \frac{R_z}{qj^{\frac{1}{q}}}\Gamma(\frac{1}{q}) (1  + j\ \alpha_0 )
	\label{eq1521} 
\end{eqnarray} 
Using  Eqs.(\ref{eq1531}), (\ref{eq153}) and (\ref{eq1521}) in Eq.(\ref{eq152}) we have,
\begin{eqnarray}
 q_{th} ({\bf  r}) &=&  \frac{1}{\lambda_{th}^3}\sum_{j=1}^\infty \frac{8 }{j^{2}}  \frac{R_x(T)R_y(T)R_z(T)\ \Gamma(\frac{1}{p}) \Gamma(\frac{1}{l}) \Gamma(\frac{1}{q})}{p l q\ j^{\eta} }   \nonumber\\
&&\ \ \ \ \ \ \ \ \ \times\ \ {(1  + j\ \alpha_0 )}  
	\label{eq154} 
\end{eqnarray}   
where
$\eta = \frac{1}{p} + \frac{1}{l} +  \frac{1}{q} + \frac{1}{2}$, 
Gathering Eq's(\ref{eq154}) and (\ref{eq4-1}) leads to,
\begin{eqnarray}
q  &=& q_0 + \frac{8}{p l q} \frac{abc  }{\epsilon_1^{1/p}\epsilon_2^{1/l}\epsilon_3^{1/q}} \Big(\frac{m}{2\pi \hbar^2}\Big)^{3/2}\ (k_B T)^{\eta+1}  \nonumber\\
 &\times& \ \Gamma(\frac{1}{p}) \Gamma(\frac{1}{l}) \Gamma(\frac{1}{q})  \Big[ \zeta(2 +\eta) + \alpha_0 \zeta(1+\eta) \Big]
\label{eqhf}
\end{eqnarray}
where $\zeta$ is the Riemann zeta function.

Theoretically the above mentioned thermodynamical parameters  results can be applied to a perfect 3D box. 
The accurate perfect 3D box trap is characterized by  $(p = l = q) = \infty,\ abc = {\cal V}/8$ where ${\cal V}$ is the volume of the box. It is clear that this result is independent on the values of $\ \epsilon_1, \epsilon_2 = $ and $\epsilon_3$. For imperfect 3D box, Hadzibabic  and co-worker \cite{Hadzibabic} have achieved an imperfect    box trap characterized by  $p,l,q > 10$. Moreover  $p,l,q  \sim $100 can be reached.

%%%%%%%%%%%%%%%%%%%%%%%%%%%%%%%%%%%%%%%%%%%%%%%%%
%%%%%%%%%%%%%%%%%%%%%%%%%%%%%%%%%%%%%%%%%%%%%%%%%%%%%%%%%%%%
\section{Condensate fraction and critical points of trapped boson atoms in imperfect 3D box } 
The phenomenon of BEC for trapped boson gas in an imperfect box  is fully described by Eq.'s (\ref{eqhf}).
Using the same procedure, one can  also obtain results for the total number of particles $N$ \cite{has0,has1},
 \begin{eqnarray}
N &=& N_0 + \frac{8}{p l q}  \frac{abc  }{\epsilon_1^{1/p}\epsilon_2^{1/l}\epsilon_3^{1/q}} \Big(\frac{m}{2\pi \hbar^2}\Big)^{3/2}\ (k_B T)^{1+ \eta}  \nonumber\\
 &\times& \ \Gamma(\frac{1}{p}) \Gamma(\frac{1}{l}) \Gamma(\frac{1}{q})  \Big[ \zeta(1 +\eta) + \alpha_0 \zeta(\eta) \Big] 
\label{con}
\end{eqnarray} 
 The condensate fraction and the critical temperature are given by  
\begin{eqnarray}
\frac{N_0}{N} &=& 1 - \frac{8}{p l q}  \frac{abc  }{\epsilon_1^{1/p}\epsilon_2^{1/l}\epsilon_3^{1/q}} \Big(\frac{m}{2\pi \hbar^2}\Big)^{3/2}\frac{(k_B T)^{1+\eta}}{N}  \nonumber\\
 &\times& \ \Gamma(\frac{1}{p}) \Gamma(\frac{1}{l}) \Gamma(\frac{1}{q})  \Big[ \zeta(1+\eta) + \alpha_0 \zeta(\eta) \Big] 
\label{eq101}
\end{eqnarray} 
the last term in Eq.(\ref{eq101}) gives the exact  the correction of the ideal gas result due to the interatomic interaction. Setting $\alpha_0 = 0$ in Eq.(\ref{eq101}) we are able to calculate the condensation temperature $T_0$ in ideal case. Since $N_0 = 0$  at $T = T_0$, then
\begin{eqnarray}
	T_0 &=& \frac{1}{k_B} \Big[ \frac{p l q}{8}  \frac{\epsilon_1^{1/p}\epsilon_2^{1/l}\epsilon_3^{1/q}}{abc  } \Big(\frac{2\pi \hbar^2}{m}\Big)^{3/2}\frac{N}{ \zeta(1+\eta) }  \nonumber\\
 &\times& \frac{1}{ \Gamma(\frac{1}{p}) \Gamma(\frac{1}{l}) \Gamma(\frac{1}{q}) } \Big]^{\frac{1}{1+\eta}} 
\end{eqnarray}
 in terms of $T_0$ Eq.(\ref{eq101}) becomes  
\begin{eqnarray}
\frac{N_0}{N} &=& 1 - \Big(\frac{T}{T_0}\Big)^{1+\eta}  \Big[ 1 + \alpha\ \Big(\frac{T_0}{T}\Big) \frac{\zeta(\eta)}{\zeta(1+\eta)} \Big] 
\label{eq102}
\end{eqnarray} 
where $\alpha = \frac{\mu}{k_BT_0}$.

\begin{figure}
\resizebox{0.50\textwidth}{!}{\includegraphics{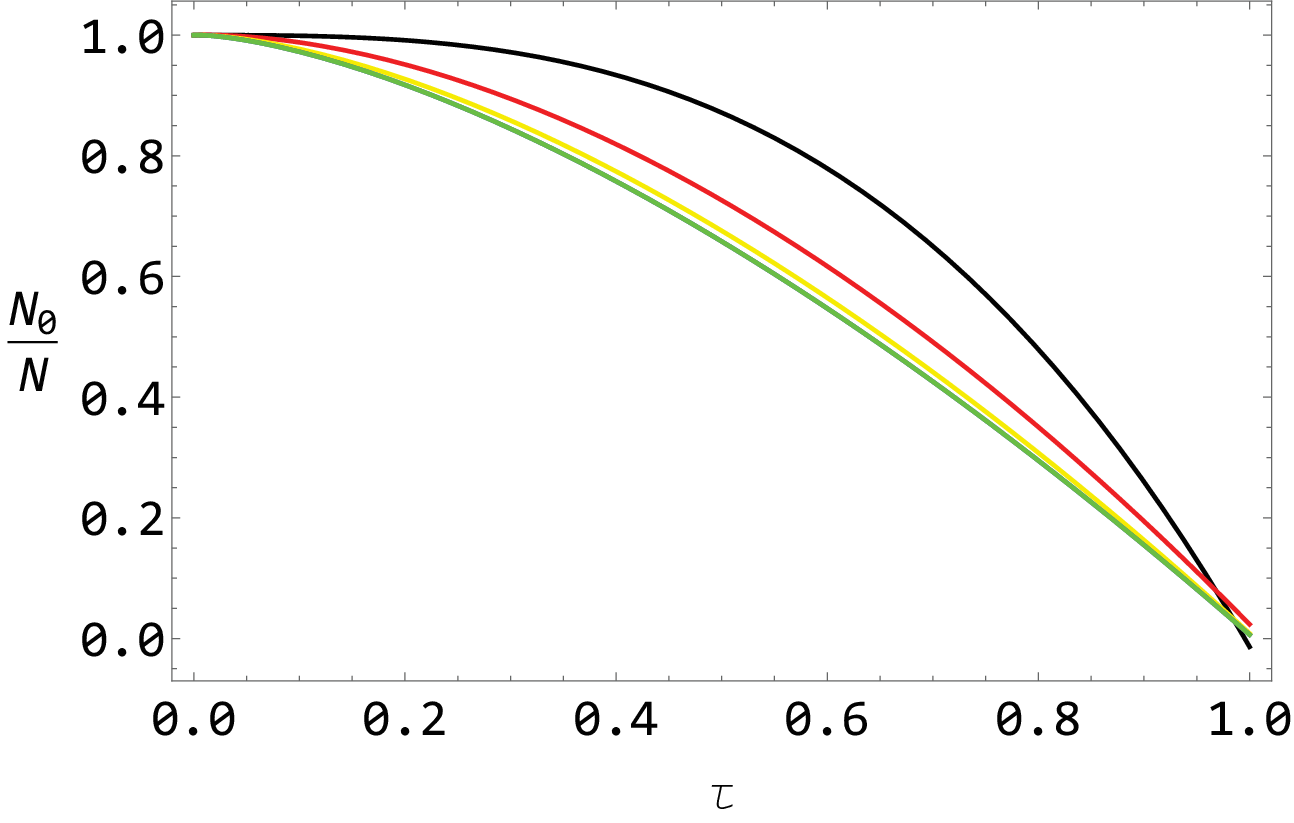}}
\caption{ Condensate fraction versus the  reduced temperature  for interacting ($\alpha = 0.01$) condensate boson atoms trapped in different  boxes: black line corresponds to $\eta$ = 2 ($p=l=q=2$ (harmonic oscillator  trap); red line corresponds to $\eta$ = 0.8 ($p=l=q=10$ (imperfect box  trap); yellow line
corresponds to $\eta$ = 0.6 ($p=l=q=30$  (imperfect box  trap); blue line corresponds to $\eta$ = 0.56 ($p=l=q=50$ (nearly perfect box  trap); green line corresponds to $\eta$ = 0.53 ($p=l=q=1000$ (close to perfect box  trap). }    
\label{f1}
\end{figure}

The calculated results from Eq.(\ref{eq102}) are represented graphically  
in figure (\ref{f1}). The plot shows that for a given temperature, the number of particles in the condensed state
decreases as the exponent $p,q$ and $l$ grows (increases $p,q$ and $l$ leads to decreases $\eta$). The above results carefully  characterise the ideal box trap that is achieved for exponent $p,q$ and $l \ge 30$.

\begin{figure}
\resizebox{0.50\textwidth}{!}{\includegraphics{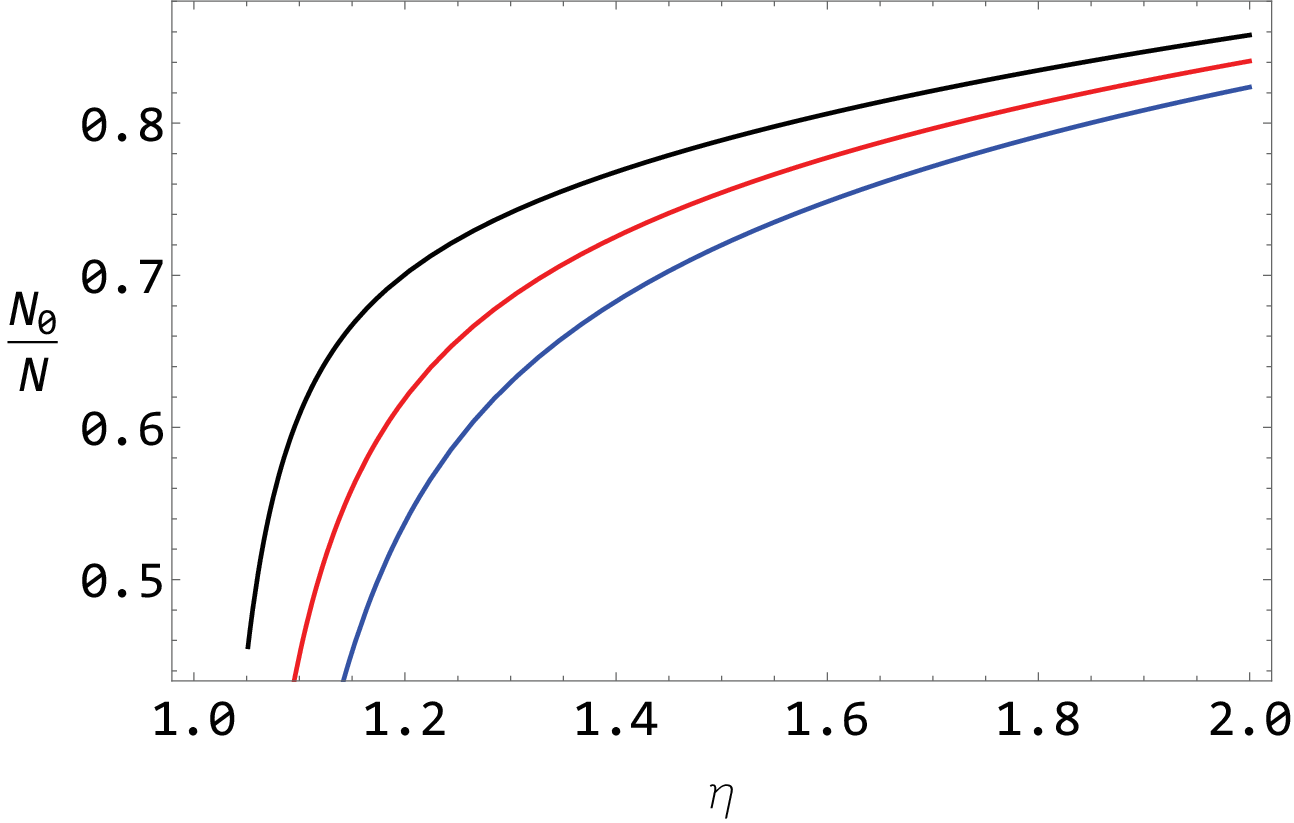}}
\caption{ Condensate fraction versus the exponent $\eta$  for different $\alpha$ = 0.05 (black line), 0.1 (red line) and 0.15 (blue line).}
\label{f2}
\end{figure}

Figure (\ref{f2}) is devoted to investigate the condensate fraction behavior as a function of the exponent $\eta$ and the interaction parameter $\alpha$.
This figure shows that the condensate fraction has a monotonically increasing nature by increasing $\eta$ until it reaches  a semi saturation values. 
The condensate fraction increases very fast at small $\eta \le 1.3$ which corresponding to $p=l=q \sim 4$. Increases the interaction parameter leads to decreases the condensate fraction as usual.

\begin{figure}
\resizebox{0.50\textwidth}{!}{\includegraphics{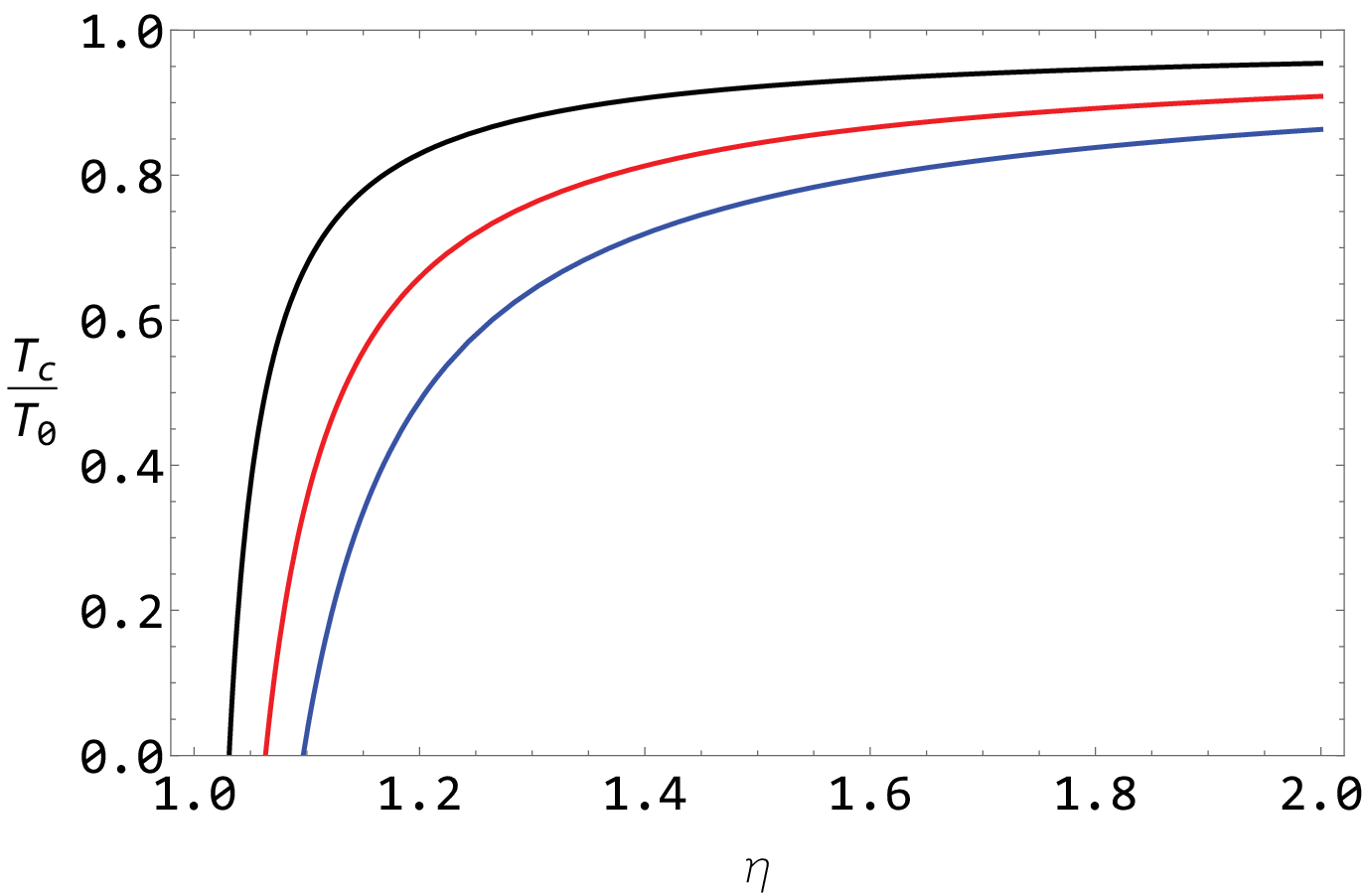}}
\caption{Critical temperature versus the exponent $\eta$  for different $\alpha$ = 0.05 (black line), 0.1 (red line) and 0.15 (blue line). }
\label{f3}
\end{figure}

One of the main goal of this work is to study the effect of  the external potential exponents $p,l$ and $q$ on the  critical temperature and critical points. The critical temperature   can be obtained as usual \cite{hau} by setting $N_0$ in  Eq.(\ref{eq102}) equal to zero, thus  
\begin{eqnarray}
T_c &\approx&  T_0   \Big[ 1- \alpha\ \frac{1}{1 + \eta}\frac{\zeta(\eta)}{\zeta(1+\eta)}\Big] 
	\label{eq18}
\end{eqnarray}
The calculated results from Eq.(\ref{eq18}) is represented graphically 
in figure (\ref{f3}). This figure shows  that the critical temperature increases monotonically very fast with increasing $\eta$, i.e. from imperfect boxes trap towards harmonic oscillator trap.

\begin{figure}
\resizebox{0.50\textwidth}{!}{\includegraphics{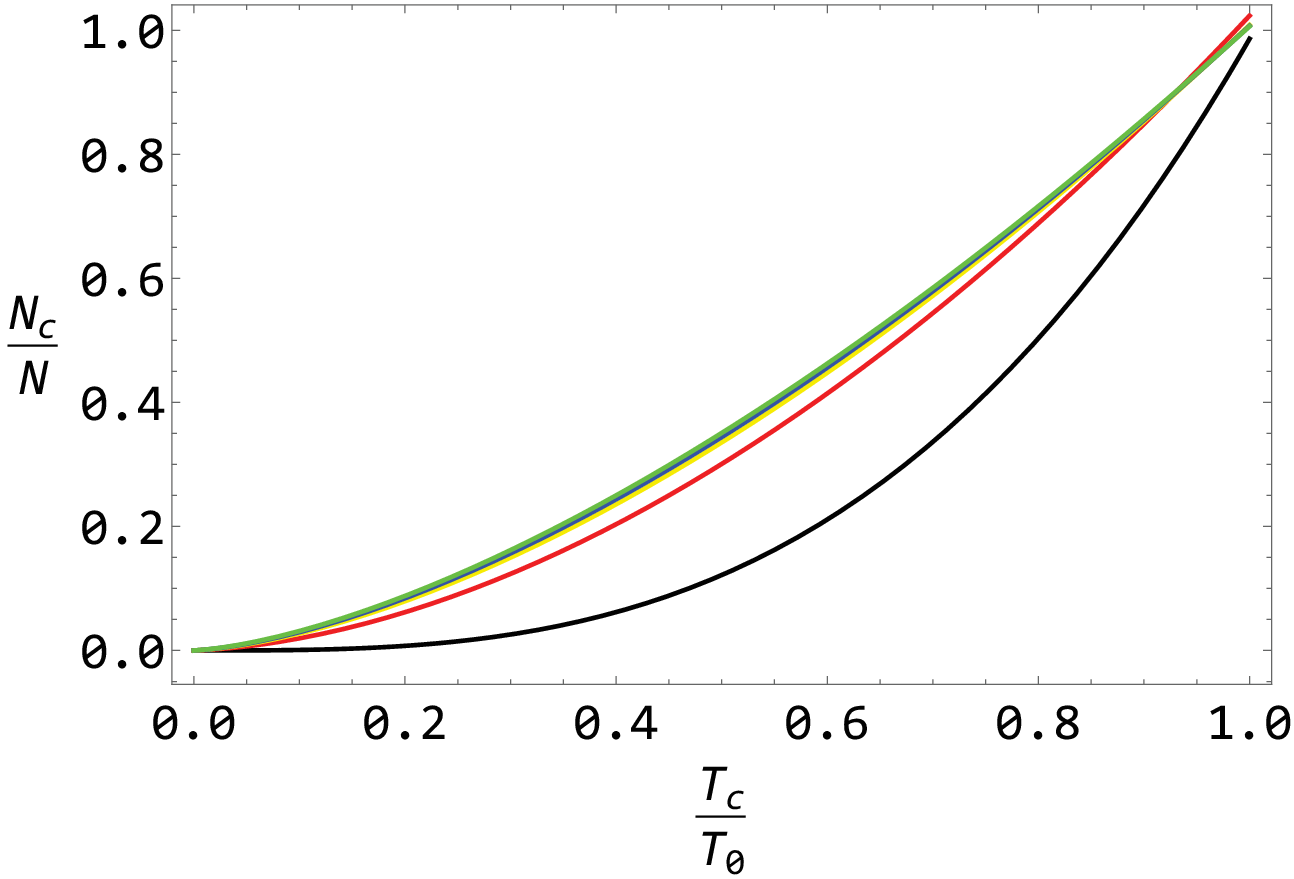}}
\caption{ Critical atoms number versus the critical temperature  for different $\alpha$ = 0.05 (black line), 0.1 (red line) and 0.15 (blue line).}
\label{f4}
\end{figure}

Now it is straightforward to calculate the typical values for the threshold atom number and its corresponding temperature  at the BEC phase transition. They are also called the critical point values $(N_c,T_c)$.  If the atom number is increased beyond $N_c$ at constant temperature, or  the temperature is reduced beyond $T_c$ for constant atom number, the critical points  define the value of the atoms number and its corresponding temperature where a bimodal distribution appeared.  In our approach resolution of these conundrums lies in the observation that the excited state occupations are independent on the condensed atoms $N_0$ at the onset of condensation, but it dependent on the critical temperature, $T_c$. In this case  the critical atom number  as a function of the critical  temperature can be obtained by using the relation
\begin{equation}
	 N_c = N_{ex}(\mu = E_0, T_c),
	\label{eq16}
\end{equation}  
where $E_0$ is the lowest energy eigen values of the Hamiltonian (\ref{eq2}).
 If we write, $ N = N_0 + N_{ex}$,
 one have
\begin{eqnarray}
 \frac{N_c}{N}   &=&  \Big(\frac{T_c}{T_0}\Big)^{1+\eta}  \Big[ 1 + \alpha\ \Big(\frac{T_0}{T_c}\Big) \frac{\zeta(\eta)}{\zeta(1+\eta)} \Big] \nonumber\\
	&=& \Big[ 1- \alpha\ \frac{1}{1 + \eta}\frac{\zeta(\eta)}{\zeta(1+\eta)}\Big]^{\eta}  \Big[  1 + \alpha\ \frac{\eta}{1 + \eta}\frac{\zeta(\eta)}{\zeta(1+\eta)} \Big] \nonumber\\
 \label{eq15}
\end{eqnarray}
with $ T_c$ is given in Eq.(\ref{eq18}). The calculated results from Eq.(\ref{eq15}) is represented graphically 
in figure (\ref{f4}). This figure shows  that the critical atoms number increases monotonically  with increasing $T_c$.

\begin{figure}
\resizebox{0.50\textwidth}{!}{\includegraphics{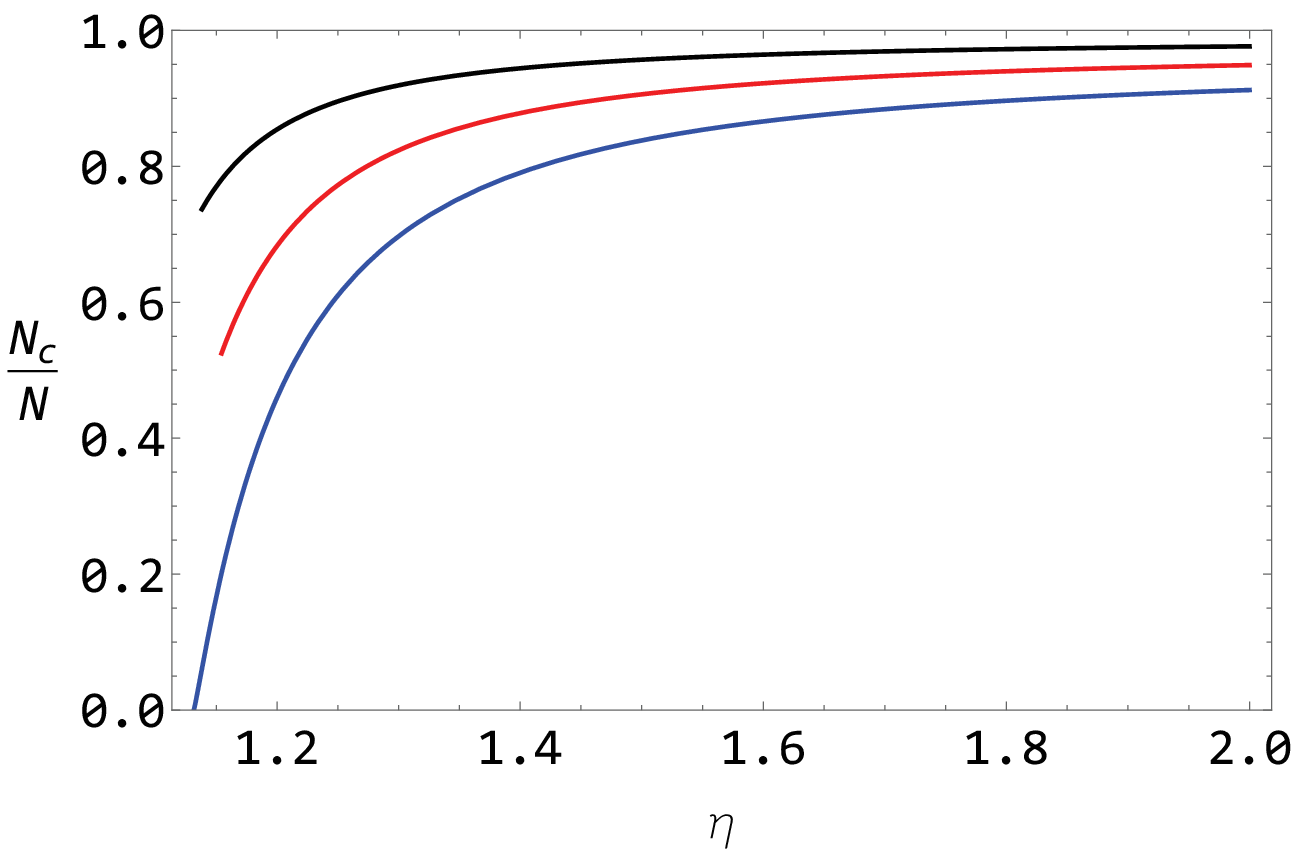}}
\caption{Critical atoms number versus the exponent $\eta$  for different $\alpha$ = 0.05 (black line), 0.1 (red line) and 0.15 (blue line).  }
\label{f5}
\end{figure}

Figure (\ref{f5}) is devoted to investigate the  behavior of the critical atoms number  as a function of the exponent $\eta$ and the interaction parameter $\alpha$.
This figure shows that the critical atoms number has a monotonically increasing nature by increasing $\eta$ until it reaches  a semi saturation values. 
The  critical atoms number increases very fast at small $\eta \le 1.3$ which corresponding to $p=l=q \sim 4$. Increasing the interaction parameter leads to decrease the critical atoms number.

%==============================================
%==============================================
\section{ Conclusion} 
 In this work we have applied the mean-field, semi-
classical  to investigate  the thermodynamical properties of  interacting bosons in power-law trap potentials. 
The expressions of several thermodynamic quantities are derived
analytically and the corrections due to the interatomic interaction are determined explicitly.
We have found that for a
range of interaction strengths the behavior of the thermodynamical quantities resembles to that of non-interacting bosons. The BEC transition in the sense of macroscopic occupation of the ground state, occurs when the short-range
interparticle interactions are not too strong.
The power-law trap potentials resembles
 the same perfect box trap for power exponent greater than 30.
  Our methods are also
suitable for studies of degenerate Fermi gases and rotating 
systems and are compatible with the implementation
of 3D optical lattices.

%===================================

\end{document}